\documentclass[aps,apl,twocolumn,superscriptaddress]{revtex4-1}%preprint

\usepackage[version=3]{mhchem} % Formula subscripts using \ce{}
\usepackage{hyperref}
\usepackage{amsfonts}
\usepackage{amsmath}
\usepackage{amssymb}
\usepackage{graphicx}
\usepackage{xcolor}

\usepackage{tabulary}
\usepackage{colortbl}

\usepackage[utf8]{inputenc}
\usepackage[T1]{fontenc}

\usepackage{color}
\usepackage{listings}
\usepackage{xpatch}

\usepackage{array}
\begin{document}

\title{Temperature-Resilient True Random Number Generation with Stochastic Actuated Magnetic Tunnel Junction Devices}

\author{Laura Rehm}
\email{laura.rehm@mail.com}
\affiliation{Center for Quantum Phenomena, Department of Physics, New York University, New York, NY 10003, USA}
\author{Md Golam Morshed}
\affiliation{Department of Electrical and Computer Engineering, University of Virginia, Charlottesville, VA 22904, USA}
\author{Shashank Misra}
\affiliation{Sandia National Laboratories, Albuquerque, New Mexico 87185, USA}
\author{Ankit Shukla}%
\affiliation{Department of Electrical and Computer Engineering, University of Illinois at Urbana-Champaign, Urbana, IL 61801, USA}
\author{Shaloo Rakheja}%
\affiliation{Department of Electrical and Computer Engineering, University of Illinois at Urbana-Champaign, Urbana, IL 61801, USA}
\author{Mustafa Pinarbasi}
\affiliation{Spin Memory Inc., Fremont, California 94538, USA}
\author{Avik W. Ghosh}%
\affiliation{Department of Electrical and Computer Engineering, University of Virginia, Charlottesville, VA 22904, USA}
\author{Andrew D. Kent}
\email{andy.kent@nyu.edu}
\affiliation{Center for Quantum Phenomena, Department of Physics, New York University, New York, NY 10003, USA}

\date{\today}

\begin{abstract} 
Nanoscale magnetic tunnel junction (MTJ) devices can efficiently convert thermal energy in the environment into random bitstreams for computational modeling and cryptography. We recently showed that perpendicular MTJs activated by nanosecond pulses can generate true random numbers at high data rates. Here, we explore the dependence of probability bias—--the deviations from equal probability (50/50) 0/1 bit outcomes—--of such devices on temperature, pulse amplitude, and duration. Our experimental results and device model demonstrate that operation with nanosecond pulses in the ballistic limit minimizes variation of probability bias with temperature to be far lower than that of devices operated with longer-duration pulses. Further, operation in the short-pulse limit reduces the bias variation with pulse amplitude while rendering the device more sensitive to pulse duration. These results are significant for designing TRNG MTJ circuits and establishing operating conditions.
\end{abstract}

\maketitle
%\section{Introduction}
Random bits are a resource needed for cryptography~\cite{Sunar2009}, hardware security~\cite{Acosta2017}, Monte Carlo simulations~\cite{Bauke2007}, and probabilistic computing~\cite{Camsari2017,Misra2023}. Present-day computations often employ pseudorandom numbers generated from a seed~\cite{DSouza1998, Fernandez1999} or use true random bits produced by relatively large-scale complementary metal-oxide semiconductor (CMOS) circuits~\cite{Holman1997,dichtl2000, stojanovski2001, Wang2019}. The demand for more efficient true random number generators (TRNGs) has driven innovation in nanoscale devices that harness physically derived stochasticity to produce random bits. An exciting development is the application of nanoscale magnetic tunnel junctions (MTJs) for this application~\cite{fu2021, fukushima2021, Rangarajan2017,kim2015}. These devices can function as 2-state elements in which one of the magnetic electrodes of the MTJ, denoted the free layer, fluctuates between up and down magnetization states by thermal activation over an energy barrier. The MTJ tunnel magnetoresistance enables reading out the magnetic state electrically. Several TRNG MTJ device concepts rely on exploiting the natural thermal fluctuations that follow an Arrhenius law~\cite{Vodenicarevic2017, Hayakawa2021, Safranski2021, Schnitzspan2023}. Their fluctuation rate is thus very (exponentially) sensitive to the temperature, device geometry, and material parameters. This creates serious obstacles to applications.

Recently, we have investigated a MTJ device in which a current pulse generates a random bit, much like a coin flip randomly generates a heads or tails outcome~\cite{Rehm2023}. We denoted our device a stochastic magnetic activated random transducer MTJ (SMART-MTJ), as pulse activation plays an essential role in the device's performance. The advantages of such a device are its speed of operation (set by the pulse sequences), and that it employs perpendicularly magnetized MTJ (pMTJs) similar to those in present-day commercial spin-transfer magnetic random access memory (MRAM) devices~\cite{Kent2015}. A major benefit of SMART-MTJ devices is that their characteristics are far less sensitive to the device
parameters and, as we demonstrate in this article, environment temperature.

In the short-pulse limit, the probabilistic behavior of the SMART device comes from the thermal distribution of the initial magnetization state that is Boltzmann distributed~\cite{Rehm2023,Shukla2023}.  
Here, we experimentally investigate and analyze the sensitivity of the probabilistic behavior of medium energy barrier pMTJs operated at different write pulse conditions by exploring small variations in the applied bath temperature, pulse amplitude, and pulse duration. 

\begin{figure*}[t]
\includegraphics[width=0.98\textwidth,keepaspectratio]{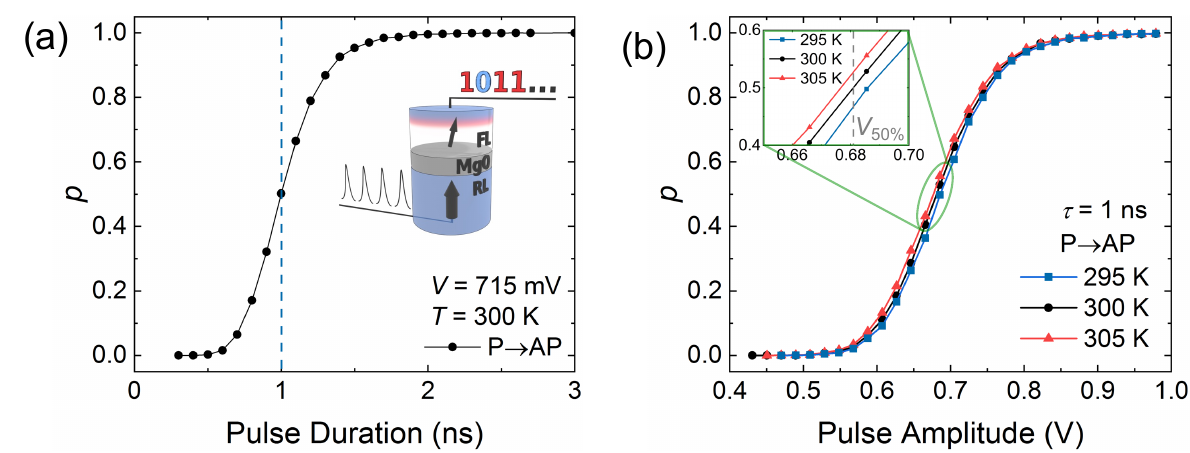}  
\caption{SMART device characteristics of the P$\rightarrow$AP transition of a 40~nm diameter pMTJ. {\bf (a)} Switching probability $p$ as a function of the pulse duration $\tau$ at room temperature with a pulse amplitude of 715~mV. SMART devices are operated at $p \approx 0.5$ and at $\tau=1$~ns, indicated by the blue dashed vertical line. The inset shows a schematic of the device concept, including the main layer structure of the MTJ with the free layer (FL), the MgO tunnel barrier , and the reference layer (RL). {\bf (b)} Switching probability $p$ as a function of the pulse amplitude $V$ at a bath temperature of $T=$~295, 300, and 305~K with a fixed pulse duration of 1~ns. The inset shows the temperature variation of the switching probability near $p = 0.5$. Each point in the plots represents 10,000 switching attempts.}
\label{Fig:1}
\end{figure*}

%\section{Perpendicular Magnetic Tunnel Junction}
The probability bias sensitivity is studied on SMART devices that are circularly shaped pMTJs with a medium energy barrier height~\cite{Rehm2023}. Specifically, the studied pMTJs are 40~nm in diameter with a room-temperature thermal stability factor $\Delta$ of 26 for the AP$\rightarrow$P transition and 51 for the P$\rightarrow$AP transition, where $\Delta=E_B/kT$, with $E_B$ the energy barrier, $k$ Boltzmann's constant, and $T$ the temperature. Our MTJs have a resistance-area product of $\simeq 3~\Omega \mu$m\textsuperscript{2}. The essential components of the SMART device are a composite CoFeB/W/CoFeB free layer stack and a CoFeB reference layer separated by a 1~nm-thin MgO tunnel barrier. The reference layer is also ferromagnetically coupled to a synthetic antiferromagnet layer structure which enables zero-field operation. A schematic of the device concept can be seen in Fig.~\ref{Fig:1}(a) and a more detailed description of the device layer structure can be found in Refs.~\cite{Rehm2019,Rehm2021}.

%\section{Experimental precedure}
The switching probability of the SMART device is explored by repeatedly applying a write-read-reset scheme. In this study, the pulse conditions for the write pulse amplitude are varied to investigate the switching probability for pulse durations between 500~ps all the way up to 100~$\mu$s. The state of the junction (either P or AP) is then read during a 150~$\mu$s-long pulse with pulse amplitudes well below the switching voltage ($V<0.03$~V). The reset pulse returns the device to a known state (\textit{e.g.}, AP state) using a 50~$\mu$s-long pulse with amplitudes well above the switching voltage. The reset and read pulses are provided from a data acquisition (DAQ) board (National Instruments PCIe-6353). The write pulses with a pulse duration between 500~ps and 100~ns are provided by an arbitrary waveform generator (Tektronix AWG 7102) while the 10 and 100~$\mu$s-long write pulses are provided by the same DAQ board.

%\section{Experimental Results and Analysis}
Figure~\ref{Fig:1}(a) shows the switching probability $p$ as a function of the applied write pulse duration $\tau$ at a fixed pulse amplitude $V =$ 715~mV and an ambient temperature of $T=$ 300~K. Each point in the graph represents $N_T = 10,000$ switching attempts. As one would expect, at a fixed pulse amplitude, the switching probability increases monotonically with pulse duration, that is, longer pulses yield a higher $p$. Now to operate the SMART device as a true random number generator, one would need to apply a 1~ns-long pulse of amplitude $V =$ 715~mV to obtain $p=0.5$ (dashed blue vertical line in Fig.~\ref{Fig:1}(a)). 
The same switching characteristics can be obtained by varying the pulse amplitude $V$ while keeping the pulse duration constant at $\tau = 1$~ns as shown in Fig.~\ref{Fig:1}(b). Higher $V$ results in more successful reversals and, therefore, a higher $p$.

\begin{figure}%[ht]
\includegraphics[width=0.48\textwidth,keepaspectratio]{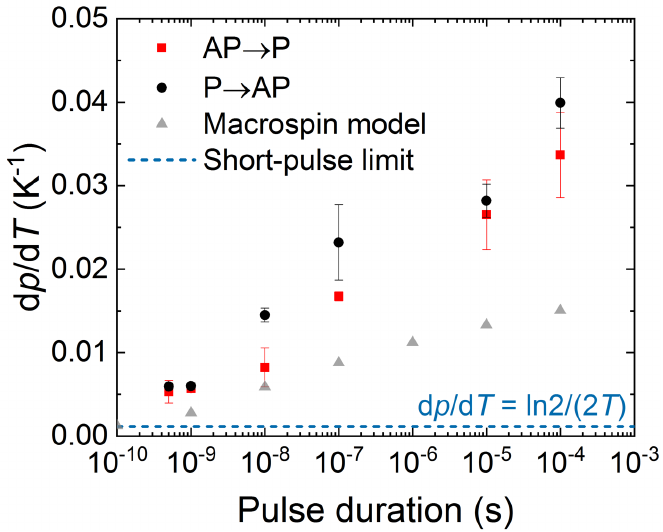}
\caption{Temperature sensitivity of the switching probability around $p = 0.5$ as a function of the applied write pulse duration at room temperature. An analysis of the short-pulse limit and a macrospin Fokker-Planck analysis show that $\mathrm{d}p/\mathrm{d}T$ has a limiting value of $\mathrm{d}p/\mathrm{d}T=\mathrm{ln}2/(2T)$ for pulse durations less than $\sim$1~ns.}
\label{Fig:2}
\end{figure}

To investigate the temperature sensitivity of the switching probability around $p = 0.5$, we set the bath temperature $T$ to 295, 300, and 305~K, and repeat the same experiment by applying write pulses with varying $V$ with $\tau = 1$~ns. We observe a shift of the curve toward lower pulse amplitudes with increasing $T$. 
If we now assume a constant $V$ with $V = V_{50\%}$, which stands for the pulse amplitude needed to reach $p = 0.5$ at $\tau = 1$~ns and $T=300$~K, we can determine the probability variation with a temperature change of $\pm5$~K. We find $p=0.47$ at $T=295$~K and $p=0.53$ at $T=305$~K (inset in Fig.~\ref{Fig:1}(b)). With this, we can deduce the temperature sensitivity of the switching probability with temperature $dp/dT$ and find a value of $dp/dT= 0.006$~K$^{-1}$ for the P$\rightarrow$AP transition. We repeat this experiment for numerous pulse durations over multiple orders of magnitude from $\tau= 500$~ps to $100\;\mu$s. Figure~\ref{Fig:2} shows the resulting $dp/dT$ as a function of the applied pulse duration $\tau$. We can clearly observe a preferable operation of the SMART device in the short-pulse limit (low-ns regime) compared to operating the same device with longer pulses. Writing the device with low-ns pulses results therefore in a lower temperature sensitivity with $dp/dT \leq 0.006$~K$^{-1}$ for $\tau = 500$~ps. In contrast, operating the device in the thermally-assisted spin-transfer switching regime ($\tau \gg 10$~ns) results in temperature sensitivities of up to $dp/dT \approx 0.04$~K$^{-1}$ for $\tau = 100\;\mu$s. 

 Our experimental results are in accord with our SMART-device model. In a macrospin model describing the ballistic switching limit, a linear approximation of the switching probability near $p \approx 0.5$ is given by~\cite{Rehm2023}:
\begin{equation}
p(V) = \frac{1}{2}+\frac{\tau \ln 2}{\tau_D V_{c0}}(V-V_{50\%}),
\label{Eq:Pswitch} 
\end{equation}
where $\tau_D$ is the intrinsic time scale for the dynamics, $V_{c0}$ is the switching threshold bias in the long pulse limit, and $V_{50\%}$ is defined as:
\begin{equation}
V_{50\%}= V_{c0} + \frac{\tau_D V_{c0}}{2\tau} \ln\left(\frac{\pi^2\Delta}{4\ln 2}\right). 
\label{Eq:V50} 
\end{equation}
Assuming a fixed pulse amplitude of $V = V_{50\%}$ and  
\begin{equation}
\frac{dV_{50\%}}{dT} = \frac{\tau_D V_{c0}}{2\tau\Delta}\frac{d\Delta}{dT}, 
\label{Eq:dV50dT} 
\end{equation}
we obtain a temperature sensitivity of the switching probability of
\begin{equation}
\frac{dp}{dT} = \frac{\ln 2}{2T}.
\label{Eq:dpdT} 
\end{equation}
Remarkably, it does not depend on any material parameters, but only on the temperature itself. At room temperature $T = 300$~K, we then find $dp/dT = 0.0016$~K$^{-1}$ from our analytic model, which can be considered the lower limit for the temperature sensitivity of the probability in the ballistic switching limit (dashed blue line in Fig.~\ref{Fig:2}). In addition to our simple analytical model, we also numerically solve the switching dynamics using a 1-D Fokker–Planck equation, which also considers single-domain or macrospin dynamics~\cite{Xie2017}. Using experimentally obtained material parameters from our device~\cite{Rehm2021}, we obtain the results shown in Fig.~\ref{Fig:2} as the gray triangles. While the experimental results are about a factor of two higher than the model, the model does capture the experimental data trend. It also saturates to the same lower limit of $dp/dT = 0.0016$~K$^{-1}$ for short pules at $\tau \leq 100$~ps we find in our analytical model. 

\begin{figure*}%[ht]
\includegraphics[width=0.98\textwidth,keepaspectratio]{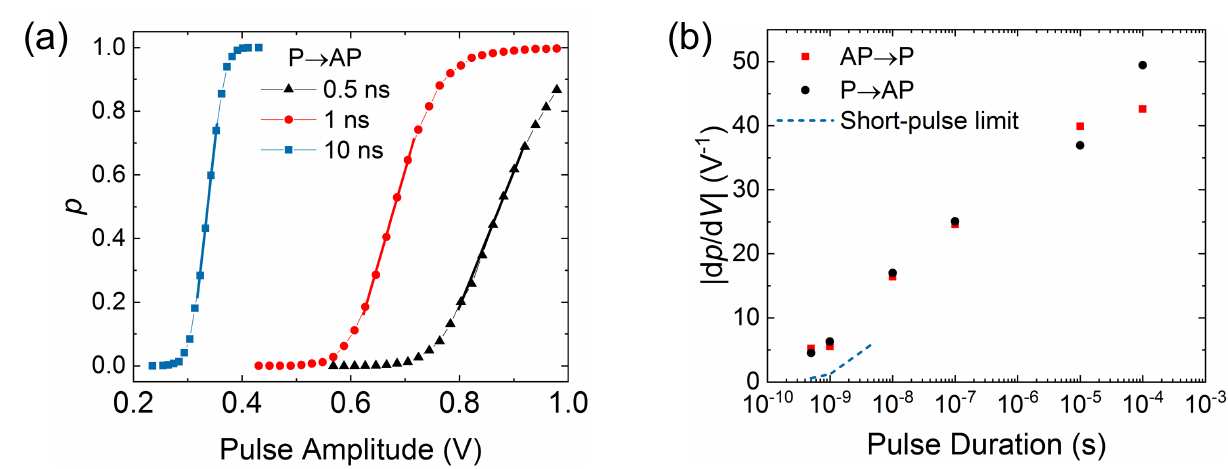}
\caption{Sensitivity of the switching probability with varying write pulse amplitude around $p = 0.5$ and at $T$ = 300~K. {\bf (a)} Switching probability versus pulse amplitude for different pulse durations of the P$\rightarrow$AP transition at room temperature. Each point is an average of 10,000 switching trials. The solid lines show the slope of the curves around $p = 0.5$. {\bf (b)} Extracted $\mathrm{d}p/\mathrm{d}V$ around $p = 0.5$ for different applied pulse durations and both switching directions. The blue dashed line represents the expectation from the short-pulse limit based on Eq.~\ref{Eq:dpdV}.}
\label{Fig:3}
\end{figure*}

In addition to $dp/dT$, we also investigated the sensitivity of $p$ to the pulse conditions. Figure~\ref{Fig:3}(a) shows $p$ as a function of the pulse amplitude $V$ for numerous $\tau$ for the P$\rightarrow$AP transition. We observe an increase in the steepness of the transition around $p = 0.5$ for increasing pulse duration. We extract the slope of the curves $dp/dV$ and the result can be seen in Fig.~\ref{Fig:3}(b). We find that $p$ is less sensitive to variation in $V$ in the ballistic limit compared to pulse durations probing the long-pulse limit. 
From Eq.~\ref{Eq:Pswitch}, we find 
\begin{equation}
\frac{dp}{dV} = \frac{\tau\ln 2}{\tau_DV_{c0}},
\label{Eq:dpdV} 
\end{equation} 
where the pulse amplitude variation of $p$ depends on material parameters through $\tau_D$ and $V_{c0}$, and is proportional to the applied pulse duration $\tau$. Using experimentally obtained material parameters to determine $\tau_DV_{c0}$ ~\cite{Rehm2021}, we can calculate the expectation from our analytical model and find that the resulting values for $\tau = 0.5-5$~ns capture the trend of the experimental data, but are underestimating the pulse amplitude sensitivity by a factor of 2 to 3 (dashed line in Fig.~\ref{Fig:3}(b)).

\begin{figure*}%[ht]
\includegraphics[width=0.98\textwidth,keepaspectratio]{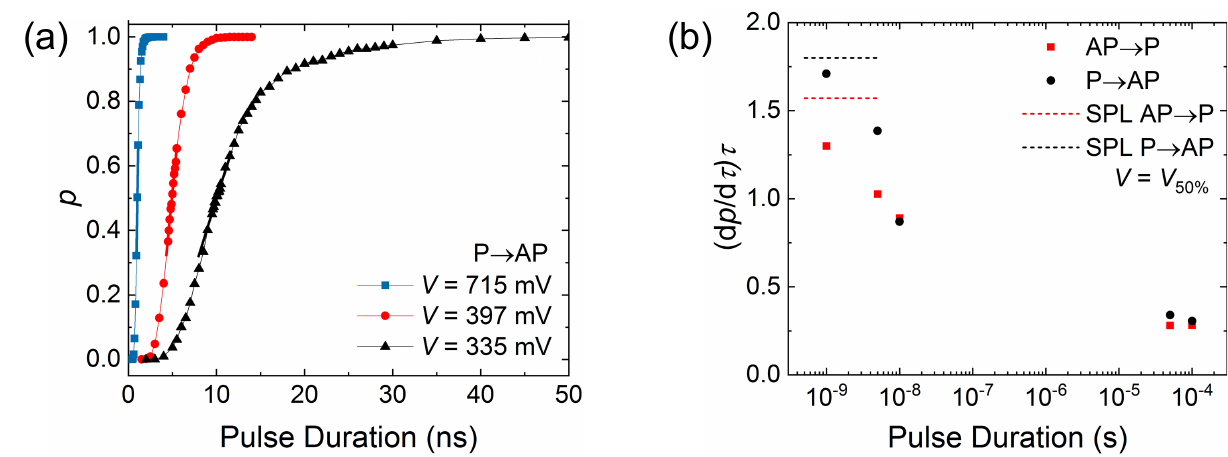}
\caption{Sensitivity of the switching probability with varying write pulse duration around $p = 0.5$. {\bf (a)} Switching probability versus pulse duration for different pulse amplitudes of the P$\rightarrow$AP transition at room temperature. Each point is an average of 10,000 switching trials. The solid lines show the slope of the cures at $p = 0.5$. {\bf (b)} Extracted $(\mathrm{d}p/\mathrm{d}\tau)\tau$ around $p = 0.5$ for different pulse durations and both switching directions. The corresponding applied pulse amplitudes are the pulse amplitudes needed to reach $p = 0.5$ at each specific pulse duration. The dashed lines show the expectation from the short-pulse limit based on Eq.~\ref{Eq:dpdtau} with $V = V_{50\%}$.}
\label{Fig:4}
\end{figure*}

The sensitivity to the pulse duration, on the other hand, shows the opposite trend. Figure~\ref{Fig:4}(a) shows $p$ as a function of the pulse duration at different $V$ for the P$\rightarrow$AP transition. The different $V$ correspond to $V_{50\%}$ for different pulse durations, \textit{e.g.}, a write pulse with $V = 335$~mV and $\tau = 10$~ns results in $p = 0.5$ (see black triangles in Fig.~\ref{Fig:4}(a)). We can clearly observe a strong $\tau$ dependence of $p$ for increasing $V$. We again extract the slope of the curves $dp/d\tau$ around $p = 0.5$ and this time multiply it by $\tau$ to obtain the relative change of $p$ with $\tau$, $(dp/d\tau)\tau$. The result can be seen in Fig.~\ref{Fig:4}(b) as a function of $\tau$. We find that SMART devices are more sensitive to variations in pulse duration in the ballistic limit than operating with longer pulses.
Comparing this again to our analytical model for $p=0.5$ ({\emph i.e.}, $V=V_{50\%}$) with Eq.~\ref{Eq:Pswitch} and Eq.~\ref{Eq:V50}, we find 
\begin{equation}
\left(\frac{dp}{d\tau}\right)\tau = \frac{\ln 2}{2}\ln \left(\frac{\pi^2 \Delta}{4\ln 2}\right).
\label{Eq:dpdtau} 
\end{equation} 
We obtain slightly higher values from our analytical model compared to our experimental data for $\tau=0.5-5$~ns (dashed lines in Fig.~\ref{Fig:4}(b)).

The macrospin model captures the experimental data trends for the variation of $p$ with temperature, voltage, and pulse duration within an order of magnitude and even better. The remaining discrepancies likely relate to the coherent reversal mechanism assumed in our single domain or macrospin model. Our SMART pMTJ devices are expected to exhibit nucleation of a sub-volume reversed domain and subsequent domain-wall-mediated reversal due to their device diameter $d = 40$~nm, which is larger than the critical dimension associated with single domain reversal ($d_\mathrm{c} < 20$~nm)~\cite{Sun2011,ChavesOFlynn2015,Statuto2021,Mohammadi2021}.

Comparing our findings with other nanomagnetic TRNG devices for which temperature-dependent data is available, we find much lower $p$ variation with respect to the temperature for our short pulse-driven SMART device. Considering experimental results only, Ref.~\cite{Fukushima2014} found a temperature variation of $p$ of 0.037~K$^{-1}$ for high-barrier pMTJs operated with longer pulses compared to our finding of only 0.006~K$^-1$ in the ballistic limit. It should be noted that easy-plane low-barrier magnets are expected to show reduced temperature sensitivity, but experimental results that explore the effect of temperature have not been reported thus far~\cite{Kaiser2019}. With the extracted $dp/dT$ of our SMART devices, we can now estimate the temperature stability required to reach a certain precision of the switching probabilities \textit{e.g.}, we need $dT\leq0.3$~K to guarantee a $dp\leq0.2\%$. While this is obtainable, a less strict requirement for the temperature stability would be needed if an exclusive or (XOR) operation is applied to bitstreams~\cite{Vatajelu2019}. While this adds overhead to the circuit, we have already shown that it may be necessary for certain applications~\cite{Rehm2023}.    

%\section{Conclusion}
To conclude, we find that our SMART devices are indeed much less sensitive to temperature compared to the same device operated in the thermally-assisted regime. Interestingly, we also find that the theoretical lower limit for the temperature sensitivity of the obtained bitstreams around $p \approx 0.5$ does not dependent on any material parameters, but only on the temperature itself. Additionally, we also investigated the sensitivity to the write pulse conditions and found a favorable behavior for the pulse amplitude, while we find the opposite behavior for the pulse duration for our short pulse-driven SMART devices. These findings are significant for designing TRNG MTJ circuits and establishing operating conditions. Furthermore, our results reinforce the potential of SMART devices as robust TRNG, especially in relation to temperature variations, which is one of the biggest challenges facing TRNG concepts that rely on thermal fluctuations.

\begin{acknowledgements}
We thank Jonathan Z. Sun at IBM Research for his comments on this manuscript. We acknowledge support from the DOE Office of Science (ASCR/BES) Microelectronics Co-Design project COINFLIPS and the Office of Naval Research (ONR) under award number N00014-23-1-2771. This work was also partly funded under the Laboratory Directed Research and Development program at Sandia National Laboratories. This paper describes objective technical results and analysis. Any subjective views or opinions that might be expressed in the paper do not necessarily represent the views of the U.S. Department of Energy or the United States Government. Sandia National Laboratories is a multimission laboratory managed and operated by National Technology \& Engineering Solutions of Sandia, LLC, a wholly owned subsidiary of Honeywell International Inc., for the U.S. Department of Energy’s National Nuclear Security Administration under contract DE-NA0003525.
\end{acknowledgements}

\section*{AUTHOR DECLARATIONS}
\section*{Conflict of Interest}
\vspace{-0.3 cm} 
\noindent
The authors have no conflicts to disclose.
\vspace{0.35 cm}
\section*{Author Contributions}
\vspace{-0.3 cm}
\noindent
\textbf{L. Rehm}: Formal analysis (lead); Investigation (lead); Methodology (lead); Writing – original draft (lead); Writing – review and editing (equal). 
\textbf{M. G. Morshed}: Formal analysis (equal); Investigation (equal); Writing – review and editing (equal).
\textbf{S. Misra}: Supervision (equal); Validation (equal). 
\textbf{A. Shukla}: Validation (equal); Writing – review and editing (equal).
\textbf{S. Rakheja}: Supervision (equal); Validation (equal). 
\textbf{M. Pinarbasi}: Validation (equal);  Resources (equal). 
\textbf{A. W. Gosh}: Supervision (equal); Validation (equal). 
\textbf{A. D. Kent}: Conceptualization (lead); Funding Acquisition (lead); Methodology (equal); Resources (equal); Supervision (equal); Validation (equal); Writing – original draft (equal); Writing – review and editing (equal).

\section*{Data Availability Statement}
The data that supports the findings of this study are available from the corresponding authors upon reasonable request.

\bibliography{SMART.bib}

\end{document}